\documentclass[10pt,twocolumn,conference]{IEEEtran}
\usepackage{dsfont}
\usepackage{setspace}
\usepackage{clrscode}
\usepackage{pict2e}
\usepackage[font=small,labelfont=small]{caption}

\usepackage{amsmath}
\usepackage{amssymb}
\usepackage[dvips]{graphicx}
\usepackage{epsfig}
\usepackage{algorithm}
\usepackage{algorithmic}
\usepackage{caption}
\usepackage{amsthm}
\usepackage{subcaption}
\usepackage{dsfont}
\usepackage[english]{babel}

\usepackage[ps2pdf,
bookmarks=false,
bookmarksnumbered=false, 
bookmarksopen=false, 
colorlinks=false]{}

\makeatletter

\newcommand{\Rmnum}[1]{\expandafter\@slowromancap\romannumeral #1@}
\makeatother

\newcommand{\LOS}{\text{LOS}}
\newcommand{\NLOS}{\text{NLOS}}
\newcommand{\SINR}{\text{SINR}}

\newcommand{\PC}{\text{P}^{\text{C}}}

\newcommand{\thh}{\text{th}}
\newcommand{\st}{\text{st}}
\newcommand{\nd}{\text{nd}}

\newcommand{\figsize}{.5}

\newtheorem{Theorem1}{Theorem}

\begin{document}


\title{Downlink Analysis in Unmanned Aerial Vehicle (UAV) Assisted Cellular Networks with Clustered Users}

\author{Esma Turgut and M. Cenk Gursoy
\\Department of Electrical
Engineering and Computer Science, Syracuse University, Syracuse, NY, 13244
\\E-mail: eturgut@syr.edu, mcgursoy@syr.edu.}

\maketitle

\begin{abstract}
The use of unmanned aerial vehicles (UAVs) operating as aerial base stations (BSs) has emerged as a promising solution especially in scenarios requiring rapid deployments (e.g., in the cases of crowded hotspots, sporting events, emergencies, natural disasters) in order to assist the ground BSs. In this paper, an analytical framework is provided to analyze the signal-to-interference-plus-noise ratio (SINR) coverage probability of unmanned aerial vehicle (UAV) assisted cellular networks with clustered user equipments (UEs). Locations of UAVs and ground BSs are modeled as Poison point processes (PPPs), and UEs are assumed to be distributed according to a Poisson cluster process (PCP) around the projections of UAVs on the ground. Initially, the complementary cumulative distribution function (CCDF) and probability density function (PDF) of path losses for both UAV and ground BS tiers are derived. Subsequently, association probabilities with each tier are obtained. SINR coverage probability is derived for the entire network using tools from stochastic geometry. Finally, area spectral efficiency (ASE) of the entire network is determined, and SINR coverage probability expression for a more general model is presented by considering that UAVs are located at different heights. Via numerical results, we have shown that UAV height and path-loss exponents play important roles on the coverage performance. Moreover, coverage probability can be improved with smaller number of UAVs, while better area spectral efficiency is achieved by employing more UAVs and having UEs more compactly clustered around the UAVs.
\end{abstract}

\begin{IEEEkeywords}
Unmanned aerial vehicle (UAV), cellular networks, SINR coverage probability, Poisson point process, Poisson cluster process, Thomas cluster process, stochastic geometry.
\end{IEEEkeywords}

\thispagestyle{empty}




\section{Introduction}
Mobile data demand has been growing exponentially in recent years due to, e.g., ever increasing use of smart phones, portable devices, and data-hungry multimedia applications. In order to meet this increasing data demand, new technologies and designs have been under consideration for fifth generation (5G) cellular networks. One of them is expected to be the deployment of dense low-power small-cell base stations (BSs) to assist the congested lower-density high-power large-cell BSs by offloading some percentage of their user equipments (UEs), resulting in a better quality of service per UE \cite{Andrews}, \cite{Elsawy}. Additionally, in the case of unexpected scenarios such as disasters, accidents, and other emergencies or temporary events requiring the excessive need for network resources such as concerts and sporting events, it is important to provide wireless connectivity rapidly \cite{Merwaday}, \cite{Bor-Yaliniz}. In such scenarios, the deployment of unmanned aerial vehicle (UAV) BSs, also known as drone BSs, has attracted considerable attention recently as a possible solution.

In \cite{Al-Hourani}, optimal altitude of low-altitude aerial platforms (LAPs) providing maximum coverage is studied. Coverage probability expression is obtained for a UAV network as a function of network and environmental parameters, and their effect on the performance is investigated in \cite{Galkin}. In \cite{Chetlur}, authors derived the coverage probability expression for a finite network of UAVs by modeling the locations of UAVs as a uniform binomial point process (BPP). Aggregate interference from neighboring UAVs and the link coverage probability are derived in \cite{Azari1} to obtain the optimum antenna beamwidth, density and altitude. In \cite{Zhang}, authors studied spectrum sharing in the deployment of aerial BSs within cellular networks and obtained the optimal drone small-cell (DSC) BS density to maximize the downlink throughput in different scenarios. An efficient 3-D placement algorithm for drone-cells in cellular networks is proposed in \cite{Bor-Yaliniz2}. In \cite{Mozaffari1}, optimal 3D deployment of multiple UAVs is also investigated to maximize the downlink coverage performance using circle packing theory. Mathematical tools of optimal transport theory is used to determine the optimal deployment and cell association of UAVs in \cite{Mozaffari2}, and the delay-optimal cell association considering both terrestrial BSs and UAVs in \cite{Mozaffari3}. Same authors have analyzed the coverage and rate performance of a network consisting of a single UAV and underlaid device-to-device (D2D) users in \cite{Mozaffari4}. In \cite{Guvenc}, performance of inter-cell interference coordination (ICIC) and cell range expansion (CRE) methods are studied for a public safety communications (PSC) heterogeneous network consisting of UAVs. Employment of emergency flexible aerial nodes is studied for the communication recovery in situations such as natural disasters in \cite{Mekikis}. Uplink performance of a two-cell cellular network with a terrestrial BS and an aerial BS is studied in \cite{Zhou} to provide better coverage probability in temporary events.

Stochastic geometry is a powerful mathematical tool to analyze the system performance of cellular networks. Hence, in most recent studies on 2D cellular networks, BS locations are assumed to follow a point process and the most commonly used distribution is the Poisson point process (PPP) due to its tractability and accuracy in approximating the actual cellular network topology \cite{Elsawy}, \cite{Elsawy2}. A similar stochastic geometry analysis can be conducted for a network of UAVs by considering UAVs distributed randomly in 3D space. Moreover, locations of the user equipments (UEs) are modeled by a Poisson cluster process (PCP) in recent studies. In \cite{Ganti}, authors analyzed the large random wireless networks by considering the locations of the nodes distributed according to a PCP on the plane. Performance of a device-to-device (D2D) network in which the device locations are modeled as a PCP is studied in \cite{Afshang1} for two realistic content availability setups. In \cite{Esma2}, the uplink performance of D2D-enabled millimeter wave (mmWave) cellular networks with clustered D2D UEs are studied. The cumulative density function (CDF) of the nearest neighbor and contact distance distributions are derived for the Thomas cluster process (TCP) in \cite{Afshang2} and for the Mat\'ern cluster process (MCP) in \cite{Afshang3} which are the special cases of PCP. In addition to modeling locations of UEs as a PCP, small-cell BS clustering is considered in \cite{Afshang4} to capture the correlation between the large-cell and small-cell BS locations. In \cite{Afshang5}, authors develop a unified heterogeneous network model in which a fraction of UEs and arbitrary number of BS tiers are modeled as PCPs to reduce the gap between the 3GPP simulation models and the popular PPP-based analytic models for heterogeneous networks. A $K$-tier heterogeneous network model in which the locations of UEs are modeled by a PCP with one small-cell BS located at the center of each cluster process is studied in \cite{Saha} for two different types of PCPs. In \cite{Sharon}, a similar heterogeneous network model with user-centric small cell deployments is developed by considering the distinguishing features of mmWave communication.

In this work, we consider a two-tier downlink network in which a network of UAVs operating at a certain altitude above ground coexisting with a network of ground BSs. Our main contributions can be summarized as follows:

\begin{itemize}
\item We provide an analytical framework to analyze the downlink coverage performance of UAV assisted cellular networks with clustered UEs by using tools from stochastic geometry. UAVs are considered to coexist with the ground BSs in the network, and locations of both UAVs and BSs are modeled as independent homogeneous PPPs. Since UAVs are planned to be deployed in overloaded scenarios, the UEs are expected to form clusters around the UAVs. Therefore, unlike previous works where the user equipment (UE) and UAV locations are assumed to be uncorrelated, we model the locations of UEs as a PCP to provide a more appropriate and realistic model.

\item CCDFs and PDFs of the path losses for each tier are derived. Then, association probabilities are obtained by considering averaged biased received
power cell association criterion. Different from \cite{Sharon} and \cite{Esma}, UAV height is taken into account in the derivation of CCDF and PDF of path losses for UAVs.

\item Laplace transforms of interferences from each tier are obtained using tools from stochastic geometry to calculate the total SINR coverage probability of the network.

\item Area spectral efficiency (ASE) of the entire network is determined. We have provided the design insights in Numerical Results section to improve network performance. In particular, we have shown that an optimal value for UAV density, maximizing the ASE, exists and this optimal value increases when UEs are located more compactly in the clusters.

\item An extension is provided to the baseline model by considering that UAVs are located at different heights. SINR coverage probability expression for this more general and practical model is presented.

\end{itemize}

The rest of the paper is organized as follows. In Section \ref{sec:system_model}, system model is introduced. Statistical characterization of the path-loss and association probabilities are also provided in Section \ref{sec:system_model}.  In Section \ref{sec:SINR Coverage Analysis}, downlink SINR coverage probability of the network is derived. ASE is formulated in Section \ref{sec:Area Spectral Efficiency}. In Section \ref{sec:Simulation and Numerical Results}, simulations and numerical results are presented to identify the impact of several system parameters on the performance metrics. Finally, conclusions and suggestions for future work are provided in Section \ref{sec:Conclusion}. Proofs are relegated to the Appendix.

\section{System Model} \label{sec:system_model}
\begin{figure}
\centering
  \includegraphics[width=\figsize\textwidth]{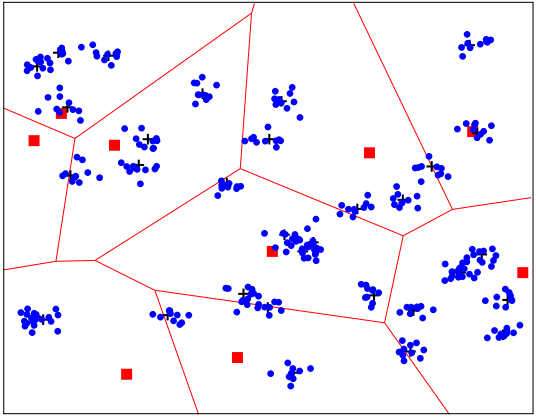}
  \caption{\small UAVs (black plus signs) and BSs (red squares) are distributed as independent PPPs, UEs (blue dots) are normally distributed around projections of UAVs on the ground.   \normalsize}
\label{Fig_Network_Model}
\end{figure}

In this section, the system model for UAV assisted cellular networks with clustered UEs is presented. We consider a two-tier downlink network, where the UAVs and ground BSs are spatially distributed according to two independent homogeneous PPPs $\Phi_{U}$ and $\Phi_{B}$ with densities $\lambda_{U}$ and $\lambda_{B}$, respectively, on the Euclidean plane. UAVs are placed at a height of $H$ above the ground, and $H$ is assumed to be constant\footnote{Subsequently, extension to considering multiple height values is also addressed.}. UAVs are deployed to provide relief to the ground cellular BSs by offloading traffic from them around hotspots or large gatherings such as sporting events or concerts. They can also be deployed during emergencies or other instances during which ground BS resources are strained \cite{Guvenc}. UEs are clustered around the projections of UAVs on the ground, and the union of cluster members' locations form a PCP, denoted by $\Phi_C$. Since UEs are located in high UE density areas, they are expected to be closer to each other forming clusters. Therefore, PCP is a more appropriate and accurate model than a homogeneous PPP. In this paper, we model $\Phi_C$ as a Thomas cluster process, where the UEs are symmetrically independently and identically distributed (i.i.d.) around the cluster centers, (which are projections of UAVs on the ground), according to a Gaussian distribution with zero mean and variance $\sigma_c^2$, and the probability density function (PDF) and complementary cumulative distribution function (CCDF) of a UE's location are given, respectively, by \cite{Haenggi}
\begin{align}
f_{D}(d)&=\frac{d}{\sigma_c^2} \exp\left( -\frac{d^2 }{2\sigma_c^2}\right), \quad d \in \mathbb{R}^2, \nonumber \\
\bar{F}_{D}(d)&=\exp\left( -\frac{d^2 }{2\sigma_c^2}\right), \quad d \in \mathbb{R}^2. \label{Fbar_R}
\end{align}
Without loss of generality, a typical UE is assumed to be located at the origin according to Slivnyak's theorem \cite{Baccelli}, and it is associated with the tier providing the maximum average biased-received power. Also, we consider an additional tier, named as $0^{\thh}$ tier that only includes the UAV at the cluster center of the typical UE similarly as in \cite{Saha} and \cite{Sharon}. Thus, our model consists of three tiers; a $0^{\thh}$ tier cluster-center UAV, $1^{\st}$ tier UAVs, and $2^{\nd}$ tier ground BSs. The proposed network model is shown in Fig. \ref{Fig_Network_Model}.

Link between a UAV and the typical UE can be either a line-of-sight (LOS) or non-line-of-sight (NLOS) link. Path-loss in NLOS links is generally higher than the path-loss in LOS links due to the reflection and scattering of signals. Therefore, an additional path-loss is experienced in NLOS links. Specifically, the path-loss of LOS and NLOS links in tier $k$ for $k=0,1$ can be modelled as follows \cite{Al-Hourani}, \cite{Al-Hourani2}:
\begin{align}\label{PL_model}
L_{k,\LOS}(r)&= \eta_{\LOS}(r^2+H^2)^{\alpha_{\LOS}/2}  \nonumber \\
L_{k,\NLOS}(r)&= \eta_{\NLOS}(r^2+H^2)^{\alpha_{\NLOS}/2}
\end{align}
where $r$ is the distance between the typical UE and the cluster center of the UAVs on the 2-D plane, i.e., projections of UAVs on the ground, $H$ is the UAV height, $\alpha_{\LOS}$ and $\alpha_{\NLOS}$ are the path-loss exponents, $\eta_{\LOS}$  and $\eta_{\NLOS}$ are the additional path losses in LOS and NLOS links, respectively. Path-loss for the $2^{\nd}$ tier ground BSs can be modeled by $L_{2}(r)=\eta_B r^{\alpha_B}$ where $\eta_B$ is the additional path-loss over the free space path-loss and $\alpha_B$ is the path-loss exponent. Similar to the UAV-to-typical UE link, the link between a BS and the typical UE can have two states, namely LOS and NLOS, with a LOS probability function which depends on the size and the density of the blockages in the environment. When communication occurs in mmWave frequency bands, the effect of blockages plays an important role and cause a significant difference between the LOS and NLOS path losses in the BS-to-typical UE link. Although the analysis of two-state path-loss model would be very similar to that of the UAV-to-typical UE link, in this paper, we consider the transmission in lower frequencies in which the difference between the LOS and NLOS path losses is not very large, and we model the path-loss in the link between the BS and the typical UE using a single state. Regarding the probability of LOS in UAV links, different models have been proposed in the literature. In this paper, we adopt the model proposed in \cite{Al-Hourani}:
\begin{equation}
\mathcal{P}_{\LOS}(r)=\frac{1}{1+b\exp\left(-c\left(\frac{180}{\pi}\tan^{-1}\left(\frac{H}{r}\right)-b\right)\right)} \label{LOS_probability}
\end{equation}
where $b$ and $c$ are constants which depend on the environment. As can be seen in (\ref{LOS_probability}), probability of having a LOS connection increases as the height of the UAVs increases.

\subsection{Statistical Characterization of the Path Loss}
We first characterize the complementary cumulative distribution function (CCDF) and the probability density function (PDF) of the path-loss in the following lemmas and corollaries.

\emph{Lemma 1:} The CCDF of the path-loss from the typical UE to a $0^{\thh}$ tier UAV can be formulated as
\begin{align}
& \bar{F}_{L_{0}}(x)= \hspace{-0.5cm} \sum_{s \in \{\LOS,\NLOS\}} \hspace{-0.5cm} \mathcal{P}_{s}(r) \bar{F}_{L_{0,s}}(x) \nonumber \\
&= \hspace{-0.5cm}\sum_{s \in \{\LOS,\NLOS\}} \hspace{-0.5cm} \mathcal{P}_{s}\left(\left(\frac{x}{\eta_s}\right)^{2/\alpha_s}\hspace{-0.5cm}-H^2\right)  \exp\left(-\frac{1}{2\sigma_c^2} \left(\left(\frac{x}{\eta_s}\right)^{2/\alpha_s}\hspace{-0.5cm}-H^2\right)\right) \label{CCDF_0}
\end{align}
where $\mathcal{P}_{\LOS}(r)$ is given in (\ref{LOS_probability}) and $\mathcal{P}_{\NLOS}(r)=1- \mathcal{P}_{\LOS}(r)$.

\emph{Proof:} See Appendix \ref{Proof of Lemma 1}.

\emph{Lemma 2:} CCDF of the path-loss from the typical UE to a $1^{\st}$ tier UAV is given by
\begin{equation}
\bar{F}_{L_1}(x)=\exp(-\Lambda_1([0,x))) \label{CCDF_1}
\end{equation}
where $\Lambda_1([0,x))$ is defined as follows:
\begin{align}
\Lambda_1([0,x))&= \Lambda_{1,\LOS}([0,x))+\Lambda_{1,\NLOS}([0,x)) \nonumber \\
&= \hspace{-0.5cm} \sum_{s \in \{\LOS,\NLOS\}} \hspace{-0.5cm} 2\pi\lambda_U \int_{0}^{\sqrt{(x/\eta_s)^{2/\alpha_s}-H^2}} \mathcal{P}_{s}(r) r dr. \label{intensity_function_1}
\end{align}

Similarly, the CCDF of the path-loss from the typical UE to a $2^{\nd}$ tier BS is given by
\begin{align}
\bar{F}_{L_2}(x)=  \exp(-\Lambda_2([0,x))) \label{CCDF_2}
\end{align}
where $\Lambda_2([0,x))=\pi\lambda_B (x/\eta_B)^{2/\alpha_B}$.

\emph{Proof:} See Appendix \ref{Proof of Lemma 2}.

\emph{Corollary 1:} The PDF of the path-loss from the typical UE to a LOS/NLOS $0^{\thh}$ tier UAV can be computed as
\begin{align}
f_{L_{0}}(x)=\sum_{s \in \{\LOS,\NLOS\}} \mathcal{P}_{s}(r) f_{L_{0,s}}(x)
\end{align}
where $f_{L_{0,s}}(x)$ is given by
\begin{align}
f_{L_{0,s}}(x)&=-\frac{d\bar{F}_{L_{0,s}}(x)}{dx}   \nonumber \\
&= \frac{1}{\sigma_c^2} \frac{x^{2/\alpha_s-1}}{\alpha_s \eta_s^{2/\alpha_s}} \exp\left(-\frac{1}{2\sigma_c^2} \left(\left(\frac{x}{\eta_s}\right)^{2/\alpha_s}\hspace{-0.5cm}-H^2\right)\right). \label{f_L0s}
\end{align}

\emph{Corollary 2:} The PDF of the path-loss from the typical UE to a LOS/NLOS $1^{\st}$ tier UAV can be computed as
\begin{equation}
f_{L_{1,s}}(x)=-\frac{d\bar{F}_{L_{1,s}}(x)}{dx}=\Lambda_{1,s}^{\prime}([0,x)) \exp(-\Lambda_{1,s}([0,x)))
\end{equation}
where $\Lambda_{1,s}^{\prime}([0,x))$ is obtained as follows using the Leibniz integral rule:
\begin{equation}
\Lambda_{1,s}^{\prime}([0,x)) =2\pi\lambda_U \frac{x^{2/\alpha_s-1}}{\alpha_s \eta_s^{2/\alpha_s}}\mathcal{P}_{s}\left(\sqrt{\left(\frac{x}{\eta_s}\right)^{2/\alpha_s}\hspace{-0.4cm}-H^2}\right). \label{Lambda_1s_prime}
\end{equation}
Similarly, the PDF of the path-loss from the typical UE to a $2^{\nd}$ tier BS is given by
\begin{equation}
f_{L_{2}}(x)=-\frac{d\bar{F}_{L_{2}}(x)}{dx}=\Lambda_{2}^{\prime}([0,x)) \exp(-\Lambda_{2}([0,x))) \label{f_l2}
\end{equation}
where $\Lambda_{2}^{\prime}([0,x))=2\pi\lambda_B \frac{x^{2/\alpha_B-1}}{\alpha_B \eta_B^{2/\alpha_B}}$.

\subsection{Cell Association}
In this work, we consider a flexible cell association scheme similarly as in \cite{Andrews2}, \cite{Esma}. In this scheme, UEs are assumed to be associated with a UAV or a BS offering the strongest long-term averaged biased-received power (ABRP). In other words, the typical UE is associated with a UAV or a BS in tier-$k$ for $k=0,1,2$ if
\begin{equation}
P_k B_k L_k(r)^{-1} \geq P_j B_j L_{min,j}(r)^{-1}, \text{for all} \; j=0,1,2, j \neq k
\end{equation}
where $P$ and $B$ denote the transmit power, and biasing factor, respectively, in the corresponding tier (indicated by the index in the subscript), $L_k(r)$ is the path-loss in the $k^{\text{th}}$ tier as formulated  in (\ref{PL_model}), and $L_{min,j}(r)$ is the minimum path-loss of the typical UE from a UAV or BS in the $j^{\thh}$ tier. In the following lemmas, we provide the association probabilities with a UAV/BS in the $k^{\thh}$ tier using the result of Lemma 1 and Corollary 1.

\emph{Lemma 3:} The probability that the typical UE is associated with a $0^{\thh}$ tier LOS/NLOS UAV is
\begin{align}
\mathcal{A}_{0,s}= \int_{\eta_s H^{\alpha_s}}^{\infty} \hspace{-0.1cm} &\mathcal{P}_{s}\left(\left(\frac{l_{0,s}}{\eta_s}\right)^{2/\alpha_s}\hspace{-0.5cm}-H^2\right) f_{L_{0,s}}(l_{0,s}) \nonumber \\
& \times e^{-\sum_{j=1}^{2} \Lambda_j\left(\left[0,\frac{P_j B_j}{P_0 B_0}l_{0,s}\right)\right)}dl_{0,s}  \label{Association_Prob0}
\end{align}
$\text{for } s \in \{\text{LOS }, \text{NLOS}\}$ where $\Lambda_1([0,x))$, $\Lambda_2([0,x))$, and $f_{L_{0,s}}(l_0)$ are given in (\ref{intensity_function_1}), (\ref{CCDF_2}), and (\ref{f_L0s}), respectively.
The probability that the typical UE is associated with a $1^{\st}$ tier LOS/NLOS UAV is
\begin{align}
\mathcal{A}_{1,s}=  \int_{\eta_s H^{\alpha_s}}^{\infty}& \Lambda_{1,s}^{\prime}([0,l_{1,s})) \bar{F}_{L_{0}}\left(\frac{P_0 B_0}{P_1 B_1}l_{1,s}\right) \nonumber \\
&\times e^{-\sum_{j=1}^{2} \Lambda_j\left(\left[0,\frac{P_j B_j}{P_1 B_1}l_{1,s}\right)\right)}dl_{1,s}  \label{Association_Prob1}
\end{align}
$\text{for } s \in \{\text{LOS }, \text{NLOS}\}$ where $\bar{F}_{L_{0}}(x)$, and $\Lambda_{1,s}^{\prime}([0,x))$ are given in (\ref{CCDF_0}) and (\ref{Lambda_1s_prime}), respectively.

The probability that the typical UE is associated with a $2^{\nd}$ tier BS is

\begin{equation}
\mathcal{A}_{2}= \hspace{-0.2cm} \int_0^{\infty} \hspace{-0.3cm} \Lambda_{2}^{\prime}([0,l_2)) \bar{F}_{L_{0}}\left(\frac{P_0 B_0}{P_2 B_2}l_2\right) e^{-\sum_{j=1}^{2} \Lambda_j\left(\left[0,\frac{P_j B_j}{P_2 B_2}l_2\right)\right)}dl_2 \label{Association_Prob2}
\end{equation}
where $\Lambda_{2}^{\prime}([0,x))$ is given in (\ref{f_l2}).

\textit{Proof}: See Appendix \ref{Proof of Lemma 3}.

\section{SINR Coverage Analysis} \label{sec:SINR Coverage Analysis}
In this section, we develop a theoretical framework to analyze the downlink SINR coverage probability for the typical UE clustered around the $0^{\thh}$ tier UAV using stochastic geometry.

\subsection{Signal-to-Interference-plus-Noise Ratio (SINR)}
The SINR experienced at the typical UE at a random distance $r$ from its associated UAV/BS in the $k^{\thh}$ tier can be written as
\begin{equation}
\SINR_k=\frac{P_k h_{k,0} L_k^{-1}(r)}{\sigma_k^2+\sum_j I_{j,k}}
\end{equation}
where
\begin{equation}
I_{j,k}=\sum_{i \in \Phi_{j}\setminus{\mathcal{E}_{k,0}}} P_j h_{j,i} L_{j,i}^{-1}(r)
\end{equation}
represents the sum of the interferences from the UAVs/BSs in the $j^{\text{th}}$ tier, $h_{k,0}$ is the small-scale fading gain from the serving BS, and $\sigma_k^2$ is the variance of the additive white Gaussian noise component. Small-scale fading gains denoted by $h$ are assumed to have an independent exponential distribution in all links. According to the cell association policy, the typical UE is associated with a BS/UAV whose path-loss is $L_k(r)$, and therefore there exists no BS/UAV within a disc of radius $\frac{P_jB_j}{P_k B_k}L_{k}(r)$ centered at the origin. This region is referred to as the exclusion disc and is denoted by $\mathcal{E}_{k,0}$. \footnote{In this paper, UAVs, BSs and UEs are assumed to have omnidirectional antennas, i.e. antennas with unit gain. However, the analysis can be extended to the case of directional antennas without much difficulty. For instance, in this case, one needs to multiply the transmit powers of the serving and interfering UAVs/BSs with the antenna gain, and update the exclusion discs for each tier by considering antenna beamwidth.}

\subsection{SINR Coverage Probability}  \label{subsec:SINRcoverage} \label{sec:SINR_Coverage_Probability}
The SINR coverage probability $\PC_k(\Gamma_k)$ is defined as the probability that the received SINR is larger than a certain threshold $\Gamma_k>0$ when the typical UE is associated with a BS/UAV from the $k^{\thh}$ tier, i.e., $\PC_k(\Gamma_k)= \mathbb{P}(\SINR_k>\Gamma_k|t=k)$ where $t$ indicates the associated tier. The total SINR coverage probability $\PC$ of the network can be computed as follows:
\begin{equation}
\PC=\sum_{k=0}^1 \sum_{s \in \{\LOS,\NLOS\}}  \left[\PC_{k,s} (\Gamma_k)\mathcal{A}_{k,s}\right]+\PC_{2}(\Gamma_2)\mathcal{A}_{2}, \label{CoverageProbability}
\end{equation}
where $\PC_{k,s}(\Gamma_k)$ is the conditional coverage probability given that the UE is associated with a $k^{\thh}$ tier LOS/NLOS UAV, $\mathcal{A}_{k,s}$ is the association probability with the $k^{\thh}$ tier for $k \in \{0,1\}$, and $\PC_{2}(\Gamma_2)$ is the conditional coverage probability given that the UE is associated with a BS in the $2^{\nd}$ tier and $\mathcal{A}_{2}$ is the association probability with the $2^{\nd}$ tier. In the following theorem, we provide the main result for the total network coverage.

\begin{Theorem1}: The total SINR coverage probability of the UAV assisted cellular networks with clustered UEs is given at the top of
the next page in (\ref{total_SINR_coverage})
\begin{figure*} \small
\begin{align}
\PC &= \sum_{s \in \{\LOS,\NLOS\}} \int_{\eta_s H^{\alpha_s}}^{\infty} e^{-\frac{\Gamma_0 l_{0,s}\sigma_0^2}{P_0}} \left(\prod_{j=1}^{2} \mathcal{L}_{I_{j,0}}\left(\frac{\Gamma_0 l_{0,s}}{P_0}\right)\right) \mathcal{P}_{s}(l_{0,s}) f_{L_{0,s}}(l_{0,s})  e^{-\sum_{j=1}^{2} \Lambda_j\left(\left[0,\frac{P_j B_j}{P_0 B_0}l_{0,s}\right)\right)}dl_{0,s} \nonumber \\
&+ \sum_{s \in \{\LOS,\NLOS\}} \int_{\eta_s H^{\alpha_s}}^{\infty} e^{-\frac{\Gamma_1 l_{1,s}\sigma_1^2}{P_1}} \left(\prod_{j=0}^{2} \mathcal{L}_{I_{j,1}}\left(\frac{\Gamma_1 l_{1,s}}{P_1}\right)\right) \Lambda_{1,s}^{\prime}([0,l_{1,s})) \bar{F}_{L_{0}}\left(\frac{P_0 B_0}{P_1 B_1}l_{1,s}\right) e^{-\sum_{j=1}^{2} \Lambda_j\left(\left[0,\frac{P_j B_j}{P_1 B_1}l_{1,s}\right)\right)}dl_{1,s} \nonumber \\
&+ \int_0^{\infty} e^{-\frac{\Gamma_2 l_{2}\sigma_2^2}{P_2}} \left(\prod_{j=0}^{2} \mathcal{L}_{I_{j,2}}\left(\frac{\Gamma_2 l_{2}}{P_2}\right)\right) \Lambda_{2}^{\prime}([0,l_2)) \bar{F}_{L_{0}}\left(\frac{P_0 B_0}{P_2 B_2}l_2\right) e^{-\sum_{j=1}^{2} \Lambda_j\left(\left[0,\frac{P_j B_j}{P_2 B_2}l_2\right)\right)}dl_2\label{total_SINR_coverage}
\end{align}
\end{figure*} \normalsize
where
\begin{equation}
\mathcal{L}_{I_{0,k}}(u)= \hspace{-0.6cm}\sum_{s^{\prime} \in \{\LOS,\NLOS\}} \int_{\mathcal{E}_{0,0}}^{\infty} \frac{1}{1+ u P_0 x^{-1}} \mathcal{P}_s(x) f_{L_{0,s^{\prime}}}(x) dx \label{LT_I0}
\end{equation}
\begin{equation}
\mathcal{L}_{I_{1,k}}(u)= \hspace{-0.6cm} \prod_{s^{\prime} \in \{\LOS,\NLOS\}} \hspace{-0.6cm} \exp\left(-\int_{\mathcal{E}_{1,0}}^{\infty} \frac{u P_1 x^{-1}}{1+u P_1 x^{-1}}  \Lambda_{1,s^{\prime}}^{\prime}(dx)\right) \label{LT_I1}
\end{equation}
\begin{equation}
\mathcal{L}_{I_{2,k}}(u)=  \exp\left(-\int_{\mathcal{E}_{2,0}}^{\infty} \frac{u P_2 x^{-1}}{1+u P_2 x^{-1}}  \Lambda_{2}^{\prime}(dx)\right). \label{LT_I2}
\end{equation}
\end{Theorem1}
\emph{Proof:} See Appendix \ref{Proof of Theorem 1}.

General sketch of the proof is as follows: First, SINR coverage probability is computed given that the typical UE is associated with a $k^{\thh}$ tier LOS/NLOS UAV or a $2^{\nd}$ tier BS. Subsequently, each of the conditional probabilities are multiplied with their corresponding association probabilities, and then they are summed up to obtain the total coverage probability of the network. In order to determine the conditional coverage probabilities, Laplace transforms of interferences from each tier are obtained using tools from stochastic geometry. We also note that although the characterization in Theorem 1 involves multiple integrals, the computation can be performed relatively easily by using numerical integration tools.
\section{Area Spectral Efficiency} \label{sec:Area Spectral Efficiency}
In Section \ref{sec:SINR Coverage Analysis}, we have analyzed the SINR coverage probability performance of a UAV assisted cellular network with clustered UEs. In this section, we consider another crucial performance metric, namely area spectral efficiency (ASE), to measure the network capacity. ASE is defined as the average number of bits transmitted per unit time per unit bandwidth per unit area. It can be mathematically defined as follows:
\begin{align}
\text{ASE}= & \bigg(\lambda_U \bigg(\sum_{k=0}^1 \sum_{s \in \{\LOS,\NLOS\}}  \left[\PC_{k,s} (\Gamma_k)\mathcal{A}_{k,s}\right]\bigg)\nonumber \\
&+\lambda_B \PC_{2}(\Gamma_2)\mathcal{A}_{2}\bigg)\log_2(1+\Gamma)   \label{eq:ASE}
\end{align}
where $\PC_{k,s}(\Gamma_k)$ is the conditional coverage probability given that the UE is associated with a $k^{th}$ tier LOS/NLOS UAV for $k \in \{0,1\}$, and $\PC_{2}(\Gamma_2)$ is the conditional coverage probability given that the UE is associated with a BS in the $2^{\nd}$ tier, $\lambda_U$ and $\lambda_B$ are the average densities of simultaneously active UAV and BS links per unit area, respectively. Note that ASE defined in (\ref{eq:ASE}) is valid for a saturated network scenario, i.e., each UAV and BS has at least one cellular UE to serve in the downlink. If the network is not saturated, the presence of inactive UAVs and BSs will lead to increased SINR (due to lower interference), and coverage probability will increase. However, ASE may be lower as a result of fewer number of active links per unit area.

\section{Extension to a Model with UAVs at Different Heights}
In the preceding analysis, we consider that UAVs are located at a height of $H$ above the ground, and $H$ is assumed to be the same for all UAVs. However, the proposed analytical framework can also be employed to analyze the coverage probability when UAV height is not fixed, i.e., UAVs are assumed to be located at different heights. In this setup, we assume that there are $M$ groups of UAVs such that the $m^{\thh}$ UAV group is located at the height level $H_m$ for $m=1,2,\ldots,M$ and UAVs at each height level can be considered as a UAV-tier distributed according to an independent homogeneous PPP with density of $\lambda_{U,m}$ and the total density is equal to $\sum_{m=1}^{M} \lambda_{U,m}=\lambda_U$. Different from the preceding analysis in which we have considered a single typical UE located at the origin and named its cluster center UAV as $0^{\thh}$ tier UAV, a separate typical UE for each UAV tier needs to be considered in the coverage probability analysis for this model with UAVs at different heights. For example, when we are analyzing the coverage probability of the network for a UE clustered around an $m^{\thh}$ tier UAV, we assume that the typical UE is located at the origin and its cluster center UAV is considered as the $0^{\thh}$ tier UAV similar to the previous model. Therefore, SINR coverage probability of the network given that the typical UE is clustered around an $m^{\thh}$ tier UAV for $m=1,2,\ldots,M$ can be computed as follows:
\begin{equation}
\PC_m\!\!=\!\!\sum_{k=0}^M \sum_{\substack{s \in \{\LOS,\\ \NLOS\}}} \hspace{-0.3cm} \left[\PC_{m,k,s} (\Gamma_k)\mathcal{A}_{m,k,s}\right]+\PC_{m,M+1}(\Gamma_{M+1})\mathcal{A}_{m,M+1}, \label{CoverageProbability_multiheight}
\end{equation}
where $\PC_{m,k,s}(\Gamma_k)$ is the conditional coverage probability given that the typical UE is clustered around an $m^{\thh}$ tier UAV and it is associated with a $k^{\thh}$ tier LOS/NLOS UAV, $\mathcal{A}_{m,k,s}$ is the association probability with a $k^{\thh}$ tier LOS/NLOS UAV, $\PC_{m,M+1}(\Gamma_{M+1})$ is the conditional coverage probability given that the typical UE is clustered around an $m^{\thh}$ tier and it is associated with a BS in the $(M+1)^{\st}$ tier, and $\mathcal{A}_{m,M+1}$ is the association probability with the $(M+1)^{\st}$ tier.

\begin{Theorem1}
SINR coverage probability of the network given that the typical UE is clustered around an $m^{\thh}$ tier UAV is given at the top of the next page in (\ref{total_SINR_coverage_multiheight})
\begin{figure*} \small
\begin{align}
&\PC_m = \sum_{s \in \{\LOS,\NLOS\}} \int_{\eta_s H_m^{\alpha_s}}^{\infty} e^{-\frac{\Gamma_0 l_{0,s}\sigma_0^2}{P_0}} \left(\prod_{j=1}^{M+1} \mathcal{L}_{I_{j,0}}\left(\frac{\Gamma_0 l_{0,s}}{P_0}\right)\right) \mathcal{P}_{s}(l_{0,s}) f_{L_{0,s}}(l_{0,s})  e^{-\sum_{j=1}^{M+1} \Lambda_j\left(\left[0,\frac{P_j B_j}{P_0 B_0}l_{0,s}\right)\right)} dl_{0,s} \nonumber \\
&+ \sum_{k=1}^{M}\sum_{s \in \{\LOS,\NLOS\}} \int_{\eta_s H_k^{\alpha_s}}^{\infty} e^{-\frac{\Gamma_k l_{k,s}\sigma_k^2}{P_k}} \left(\prod_{j=0}^{M+1} \mathcal{L}_{I_{j,k}}\left(\frac{\Gamma_k l_{k,s}}{P_k}\right)\right) \Lambda_{k,s}^{\prime}([0,l_{k,s})) \bar{F}_{L_{0}}\left(\frac{P_0 B_0}{P_k B_k}l_{k,s}\right) e^{-\sum_{j=1}^{M+1} \Lambda_j\left(\left[0,\frac{P_j B_j}{P_k B_k}l_{k,s}\right)\right)}dl_{k,s} \nonumber \\
&+\int_0^{\infty} \hspace{-0.3cm} e^{-\frac{\Gamma_{M+1} l_{M+1}\sigma_{M+1}^2}{P_{M+1}}} \left(\prod_{j=0}^{M+1} \mathcal{L}_{I_{j,M+1}}\left(\frac{\Gamma_{M+1} l_{M+1}}{P_{M+1}}\right)\right) \Lambda_{2}^{\prime}([0,l_{M+1})) \bar{F}_{L_{0}}\left(\frac{P_0 B_0}{P_{M+1} B_{M+1}}l_{M+1}\right) e^{-\sum_{j=1}^{M+1} \Lambda_j\left(\left[0,\frac{P_j B_j}{P_{M+1} B_{M+1}}l_{M+1}\right)\right)}dl_{M+1}\label{total_SINR_coverage_multiheight}
\end{align}
\end{figure*} \normalsize
\end{Theorem1}
\emph{Proof:} Derivation of $\PC_m$ follows similar steps as that of $\PC$ in (\ref{total_SINR_coverage}). In particular, Laplace transforms $\mathcal{L}_{I_{0,k}}$ and $\mathcal{L}_{I_{j,k}}$ for $j=1,2,\ldots,M$ are computed using the Laplace transform equations given in (\ref{LT_I0}) and (\ref{LT_I1}), respectively, by updating UAV height as $H_j$ and UAV density as $\lambda_j$ for $j=0,1,\ldots,M$. Similarly, $\mathcal{L}_{I_{M+1,k}}$ is computed using the Laplace transform expression given in (\ref{LT_I2}). $\Lambda_j([0,x))$ for $j=1,2,\ldots,M$ and $\Lambda_{k,s}^{\prime}([0,x))$ for $k=1,2,\ldots,M$ are computed using the equations $\Lambda_1([0,x))$ and $\Lambda_{1,s}^{\prime}([0,x))$ given in (\ref{intensity_function_1}) and (\ref{Lambda_1s_prime}), respectively, by inserting the UAV height and UAV density for each tier. Similarly, $\Lambda_{M+1}([0,x))$ and $\Lambda_{M+1}^{\prime}([0,x))$ are obtained using the equations for the $2^{\nd}$ tier BSs, $\Lambda_2([0,x))$ and $\Lambda_2^{\prime}([0,x))$, respectively. Furthermore, $\bar{F}_{L_{0}}(x)$ and $f_{L_{0,s}}(x)$ are computed using (\ref{CCDF_0}) and (\ref{f_L0s}), respectively, by denoting the UAV height as $H_m$.

\section{Simulation and Numerical Results} \label{sec:Simulation and Numerical Results}
In this section, theoretical expressions are evaluated numerically. We also provide simulation results to validate the accuracy of the proposed model for the UAV-assisted downlink cellular network with clustered UEs as well as to confirm of the analytical characterizations. In the numerical evaluations and simulations, unless stated otherwise, the parameter values listed in Table \ref{Table} are used.

\begin{table}
\small
\caption{System Parameters}
\centering
  \begin{tabular}{| p{4.2cm} | p{1.4cm}| p{1.6cm}|}
    \hline
    \textbf{Description}  & \textbf{Parameter}  & \textbf{Value}  \\ \hline
    \text{Path-loss exponents} & $\alpha_{\LOS}$, $\alpha_{\NLOS}$, $\alpha_B$  & 3, 3.5, 3.5 \\ \hline
    \text{Average additional path-loss} \text{for LOS and NLOS} & $\eta_{\LOS}$, $\eta_{\NLOS}$, $\eta_B$ & 1, 10, 1 \\ \hline
    \text{Environment dependent constants} &  $b$, $c$& $11.95$, $0.136$ \\ \hline
    \text{Height of UAVs} & $H$  & $10$m \\ \hline
    \text{Transmit power} & $P_0$, $P_1$, $P_2$ & 37dBm, 37dBm, 40dBm \\ \hline
    \text{UAV and BS densities} & $\lambda_U$, $\lambda_B$ & $10^{-4}$, $10^{-5}$ $(1/m^2)$ \\ \hline
    \text{Biasing factor, SINR threshold}, \text{noise variance} & $B_k$, $\Gamma_k$, $\sigma_k^2$ $\forall k$ & 1, 0dB, -90dBm \\ \hline
    \text{UEs distribution's variance} &  $\sigma_c^2$  & 25 \\ \hline
  \end{tabular} \label{Table}
\end{table}

\begin{figure}
\centering
  \includegraphics[width=\figsize\textwidth]{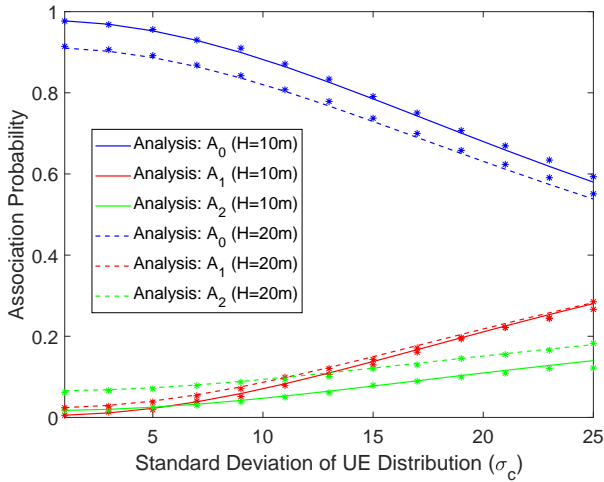}
  \caption{\small Association probability as a function of UE distribution's standard deviation $\sigma_c$ for different values of UAV height $H$. Simulation results are also plotted with markers. \normalsize}
\label{Fig_AP}
\end{figure}

First, we investigate the effect of UE distribution's standard deviation $\sigma_c$ on the association probability for different values of the UAV height $H$ in Fig. \ref{Fig_AP}. As the standard deviation increases, the UEs have a wider spread and the distances between the $0^{\thh}$ tier UAV and UEs also increase. As a result, association probability with the $0^{\thh}$ tier UAV decreases, while association probability with $1^{\st}$ tier UAVs and $2^{\nd}$ tier ground BSs increases. Similarly, $0^{\thh}$ tier association probability decreases also with the increase in the heights of the UAVs due to increase in the relative distances between the $0^{\thh}$ tier UAV and UEs. Association probability with $2^{\nd}$ tier BSs increases, while association probability with $1^{\st}$ tier UAVs remains almost unchanged. The intuitive reason behind this behavior is that when all UAVs are at a higher height, UEs are still more likely to be associated with the $0^{\thh}$ tier UAV, which is at the center of cluster, rather than $1^{\st}$ tier UAVs. Therefore, more UEs get connected to the ground BSs if the UAV height increases. Finally, we note that simulation results are also plotted in the figure with markers and there is a very good match between simulation and analytical results, further confirming our analysis.

\begin{figure}
\centering
  \includegraphics[width=\figsize\textwidth]{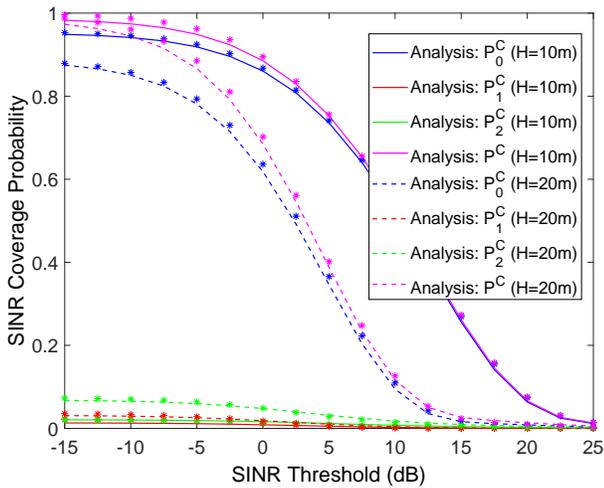}
  \caption{\small SINR Coverage probability as a function of the threshold in dB for different values of UAV height $H$. Simulation results are also plotted with markers. \normalsize}
\label{Fig_CP_differentL}
\end{figure}

Next, in Fig. \ref{Fig_CP_differentL} we plot the SINR coverage probabilities of different tiers (i.e., $P_0^C, P_1^C$ and $P_2^C$) and also the total SINR coverage probability $P^C$ as a function of the SINR threshold for different values of UAV height $H$. As seen in Fig. \ref{Fig_AP}, UEs are more likely to be associated with the $0^{\thh}$ tier UAV, which is the UAV at their cluster center, and therefore we observe in Fig. \ref{Fig_CP_differentL} that the coverage probability of $0^{\thh}$ tier UAV is much higher than that of $1^{\st}$ tier UAVs and $2^{\nd}$ tier BSs. Fig. \ref{Fig_CP_differentL} also demonstrates that the total coverage probability gets worse with the increasing UAV height as a result of the increase in the distances between the $0^{\thh}$ tier UAV and UEs. As also noted in Fig. \ref{Fig_CP_differentL}, this increase in the distances causes coverage probability of ground BSs to increase. Also similarly as before, since the association probability with the $1^{\st}$ tier UAVs remains almost unchanged with the increasing UAV height, their coverage probability also remains same.

\begin{figure}
\centering
  \includegraphics[width=\figsize\textwidth]{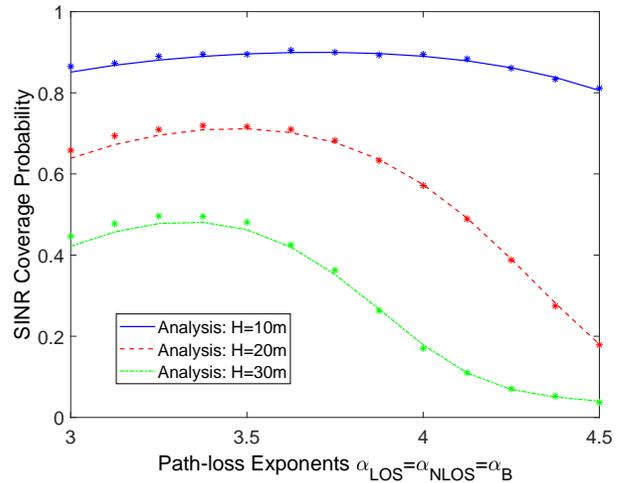}
  \caption{\small SINR coverage probability as a function of the path-loss exponents $\alpha_{\LOS}=\alpha_{\NLOS}=\alpha_B$ for different values of UAV height $H$. Simulation results are also plotted with markers. \normalsize}
\label{Fig_CP_afo_alpha}
\end{figure}

In Fig. \ref{Fig_CP_afo_alpha}, the effect of path-loss exponents on the coverage probability is investigated at different values of the UAV height by assuming $\alpha_{\LOS}=\alpha_{\NLOS}=\alpha_B$ (additional path-loss for NLOS UAV links, $\eta_{\NLOS}$, is still present.). Coverage probability initially improves when the path-loss exponents increase, but then it starts diminishing. As path-loss exponents increase, received power from the serving UAV or BS decreases, but the received power from interfering nodes also diminishes resulting in an increase in the coverage performance. However, further increasing the path-loss exponents deteriorates the coverage performance. Therefore, there exists an optimal value for path-loss exponents in which the coverage probability is maximized and this optimal value changes for different values of UAV height. For instance, we notice in the figure that the optimal value decreases when the UAV height increases. Increasing the height reduces the received power from the serving UAV, and hence lower path-loss exponent is preferred to optimize the performance. Another observation from Fig. \ref{Fig_CP_afo_alpha} is that coverage probability performance is not affected significantly from varying the path-loss exponent if the UAV height is small.

\begin{figure}
\centering
  \includegraphics[width=\figsize\textwidth]{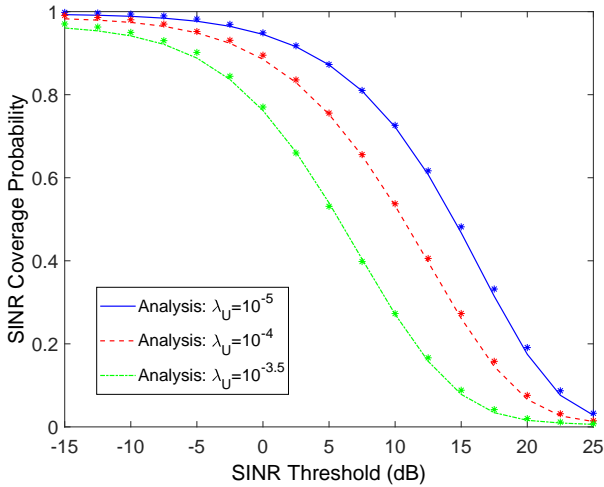}
  \caption{\small SINR coverage probability as a function of the threshold in dB for different values of UAV density $\lambda_U$. Simulation results are also plotted with markers. \normalsize}
\label{CP_different_lambdaU}
\end{figure}

\begin{figure}
\centering
  \includegraphics[width=\figsize\textwidth]{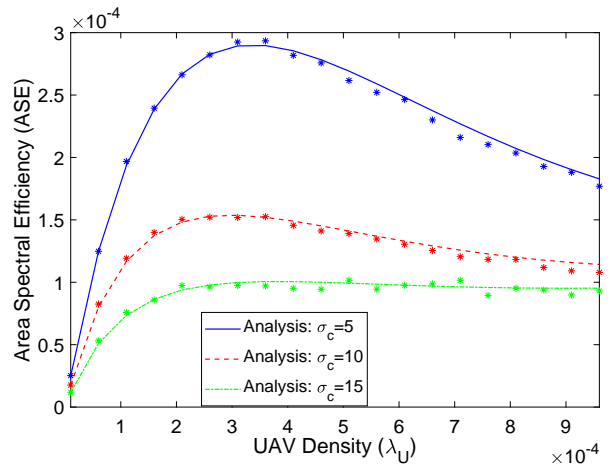}
  \caption{\small Area spectral efficiency (ASE) as a function of UAV density $\lambda_U$ for different values of UE distribution's standard deviation $\sigma_c$. Simulation results are also plotted with markers. \normalsize}
\label{ASE_different_sigmac}
\end{figure}

Next, SINR coverage probability is plotted as a function of the SINR threshold for different values of UAV density $\lambda_U$ in Fig. \ref{CP_different_lambdaU}. As shown in the figure, increase in the UAV density results in a degradation in the coverage probability. Since UEs are clustered around the projections of UAVs on the ground, they are more likely to be associated with the $0^{\thh}$ tier UAV, i.e., the UAV at their cluster center. Therefore, increasing UAV density results in higher interference levels from other UAVs and consequently lower coverage probabilities. However, as we have shown in Fig. \ref{ASE_different_sigmac} increase in UAV density leads to higher area spectral efficiency (ASE) because more UEs are covered in the network.

Specifically, in Fig. \ref{ASE_different_sigmac}, we plot ASE as a function of the UAV density $\lambda_U$ for different values of standard deviation $\sigma_c$ of the UE distribution. As the UAV density $\lambda_U$ increases, ASE first increases and then starts decreasing. This shows that there exists an optimal value for $\lambda_U$ maximizing the ASE. Below this optimal value, increasing UAV density $\lambda_U$ helps improving the spatial frequency reuse. However, after this optimal value, the effect of the increased received power from interfering UAVs offsets the benefit of covering more UEs due to having more UAVs. Furthermore, decrease in the UE distribution's standard deviation $\sigma_c$ results in a higher ASE for the same value of $\lambda_U$. Smaller $\sigma_c$ means that UEs are, on average, more compactly packed around the cluster center, and hence the distance between the UAV at the cluster center is shorter. Therefore, coverage probability is improved for smaller $\sigma_c$. Also, optimal value for $\lambda_U$ increases with decreasing $\sigma_c$ indicating that more UAVs can be deployed to support more UEs if UEs are located compactly in each cluster.

\begin{figure}
\centering
  \includegraphics[width=\figsize\textwidth]{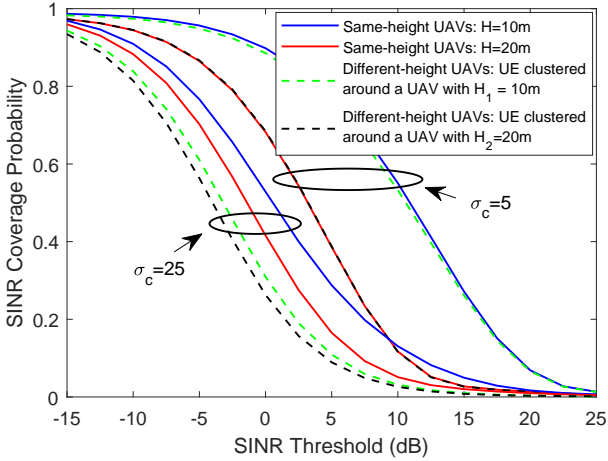}
  \caption{\small SINR coverage probability as a function of the threshold in dB for two different values of the UE distribution's standard deviation $\sigma_c$. Solid lines show the coverage probabilities when half of the UAVs are located at height $H_1 = 10$m and the other half are located at height $H_2 = 20$m, and the typical UE is clustered around a UAV at either height $H_1$ or $H_2$.  \normalsize}
\label{Fig_CP_differentL_multiheight}
\end{figure}

Finally, in Fig. \ref{Fig_CP_differentL_multiheight}, we plot the SINR coverage probability as a function of the SINR threshold for two different values of the UE distribution's standard deviation $\sigma_c$ when UAVs are assumed to be located at different heights. In this setup, we use the same parameters given in Table \ref{Table} with some differences for UAV height and UAV density. More specifically, we consider $M=2$ groups of UAVs located at altitudes $H_1=10$m and $H_2=20$m with densities $\lambda_{U,1}=\lambda_{U,2}=\lambda_U/2$ and transmit powers $P_1=P_2=37$dBm. Therefore, transmit power of the $0^{\thh}$ UAV is also equal to $P_0=37$dBm. Moreover, transmit power of the $3^{\text{rd}}$ tier ground BSs is equal to $P_3=40$dBm. In Fig. \ref{Fig_CP_differentL_multiheight}, solid lines plot the coverage probabilities when the height is the same for all UAVs. Dashed lines display the coverage probabilities when half of the UAVs are located at height $H_1$ and the other half are located at height $H_2$, and the typical UE is clustered around a UAV at either height $H_1$ or $H_2$. As shown in the figure, for $\sigma_c=5$ when the typical UE is clustered around a UAV at height $H_1=10$m in the model with two different UAV heights, it experiences almost the same coverage performance with the typical UE when all UAVs are at the same height of $H_1 = 10$m. The same observation can be made for the case of $H_2=20$m. On the other hand, when $\sigma_c$ gets larger (and hence the UEs are more widely spread around the cluster-center UAV), coverage performance in the model with UAVs at two different height levels becomes worse than that of the case in which all UAVs are at the same height. Moreover, coverage performances for the typical UEs clustered around UAVs at heights $H_1=10$m and $H_2=20$m approach each other. There are mainly three reasons behind these results: 1) association probability with the other UAVs and BSs rather than the cluster-center $0^\thh$ tier UAV increases for larger values of $\sigma_c$ (e.g., see Fig. 2); 2) when the typical UAV is clustered around a UAV at height $H_1 = 10$m, interference from half of the UAVs located at height $H_2 = 20$m is smaller than that if all UAVs were at the same height of $H_1 = 10$m, but at the same time if the UE is associated not with its cluster center UAV but with a UAV at height $H_2 = 20$m, link distance will be larger, adversely affecting the coverage probability; 3) when the typical UE is clustered around a UAV at height $H_2 = 20$m, interference from half of the UAVs located at the lower height of $H_1 = 10$m is greater but  if the UE is associated with a non-cluster-center UAV at height $H_1 = 10$m then the link quality can be better due to shorter distance. Hence, there are several interesting competing factors and tradeoffs. As a result, we observe in the case of large $\sigma_c$ that due to either increased interference or higher likelihood of being associated with a UAV at a larger height, coverage performances in the model with different UAV heights get degraded compared to the scenario in which all UAVs are at the same height.


\section{Conclusion} \label{sec:Conclusion}
In this paper, we have provided an analytical framework to compute the SINR coverage probability of UAV assisted cellular networks with clustered UEs. Moreover, we have formulated the ASE, and investigated the effect of UAV density and standard deviation of the UE distribution on the ASE. Furthermore, we have presented SINR coverage probability expression for a more general model by considering that UAVs are located at different heights. UAVs and ground BSs are assumed to be distributed according to independent PPPs, while locations of UEs are modeled as a PCP around the projections of UAVs on the ground and UEs are assumed to be connected to the tier providing the maximum average biased-received power.

Using numerical results, we have shown that standard deviation of UE distribution $\sigma_c$ and UAV height $H$ have significant impact on association probabilities. For instance, less compactly located UEs and higher UAV height lead to a decrease in the association with the cluster center UAV. We have also shown that total coverage probability can be improved by reducing the UAV height as a result of the decrease in the distances between cluster center UAV and UEs. Moreover, path-loss exponents play a crucial role in the coverage performance if the UAV height is high, and there exists an optimal value for path-loss exponents in which the coverage probability is maximized. Another important observation is that smaller number of UAVs results in a better coverage performance, while deployment of more UAVs lead to a higher ASE. Furthermore, a higher ASE can be achieved if the UES are located more compactly in each cluster. Computation of the delay-sensitive ASE (DASE) by selecting the transmission powers properly to limit the RF pollution without affecting the users' quality-of-service will be considered as future work \cite{Makki}. Furthermore, analyzing the coverage performance for a different PCP such as uniformly distributed UEs, and considering the communication in millimeter wave (mmWave) frequency bands are interesting extensions that remain as future work.

\appendix
\subsection{Proof of Lemma 1}
\label{Proof of Lemma 1}
The CCDF of the path-loss $L_{0,s}$ from the typical UE to a $0^{th}$ tier LOS/NLOS UAV can be computed as follows:
\begin{align}
\bar{F}_{L_{0,s}}(x) &= \mathbb{P}\left(L_{0,s}(r)\geq x \right) \nonumber \\
&= \mathbb{P}\left(\eta_s (d^2+H^2)^{\alpha_s/2}\geq x \right) \label{Aeq1} \\
&= \mathbb{P}\left(d \geq \sqrt{\left( \frac{x}{\eta_s}\right)^{2/\alpha_s}\hspace{-0.5cm}-H^2} \right) \nonumber \\
&= \bar{F}_{D}\left(\sqrt{\left( \frac{x}{\eta_s}\right)^{2/\alpha_s}\hspace{-0.5cm}-H^2} \right) \nonumber \\
&= \exp\left(-\frac{1}{2\sigma_c^2} \left(\left(\frac{x}{\eta_s}\right)^{2/\alpha_s}\hspace{-0.5cm}-H^2\right)\right)
\end{align}
$\text{for } s \in \{\text{LOS }, \text{NLOS}\}$ where $\bar{F}_{D}(\cdot)$ is given in (\ref{Fbar_R}) and (\ref{Aeq1}) follows from the definition of path-loss and noting that $r=d$ for $0^{\thh}$ tier. Therefore, the CCDF of the path-loss $L_{0}$ from the typical UE to a $0^{\thh}$ tier UAV can be obtained as
\begin{align}
&\bar{F}_{L_{0}}(x) =  \mathcal{P}_{\LOS}(r) \bar{F}_{L_{0,\LOS}}(x)+\mathcal{P}_{\NLOS}(r) \bar{F}_{L_{0,\NLOS}}(x) \label{Aeq2} \\
&= \hspace{-0.5cm} \sum_{s \in \{\LOS,\NLOS\}} \hspace{-0.5cm} \mathcal{P}_{s}\left(\left(\frac{x}{\eta_s}\right)^{2/\alpha_s}\hspace{-0.5cm}-H^2\right)  \exp\left(-\frac{1}{2\sigma_c^2} \left(\left(\frac{x}{\eta_s}\right)^{2/\alpha_s}\hspace{-0.5cm}-H^2\right)\right)
\end{align}
where $\mathcal{P}_{s}(r)$ is given in (\ref{LOS_probability}) and (\ref{Aeq2}) follows from the fact that there is only one UAV in the $0^{\thh}$ tier which can be a LOS or a NLOS UAV.

\subsection{Proof of Lemma 2}
\label{Proof of Lemma 2}
Intensity function for the path-loss model from the typical UE to a $1^{\st}$ tier UAV for $s \in \{\LOS,\NLOS\}$ can be computed as
\begin{align}
&\Lambda_{1,s}([0,x))=\int_{\mathbb{R}^2} \mathbb{P}\left(L_1(r)<x\right)dr \label{Aeq3} \\
&=2\pi\lambda_U \int_0^{\infty}\mathbb{P}\left(\eta_s \left(r^2+H^2\right)^{\alpha_s/2}<x\right) \mathcal{P}_{s}(r) r dr \nonumber \\
&= 2\pi\lambda_U \int_0^{\infty} \mathbb{P}\left(r< \sqrt{(x/\eta_s)^{2/\alpha_s}-H^2}\right)\mathcal{P}_{s}(r) r dr \nonumber \\
&=  2\pi\lambda_U \int_{0}^{\sqrt{(x/\eta_s)^{2/\alpha_s}-H^2}} \mathcal{P}_{s}(r) r dr \label{app:Lemma1}
\end{align}
where (\ref{Aeq3}) follows from the definition of intensity function for the point process of the path-loss. Intensity function for $2^{\nd}$ tier BSs can be also computed using the same approach. Since the link between the ground BSs and the typical UE has only one state, intensity function expression in (\ref{app:Lemma1}) reduces to $\Lambda_2([0,x))=\pi\lambda_B (x/\eta_B)^{2/\alpha_B}$.

\subsection{Proof of Lemma 3}
\label{Proof of Lemma 3}
Association probability with a $0^{\thh}$ tier LOS/NLOS UAV can be computed as follows:
\begin{align}
&\mathcal{A}_{0,s}=\mathbb{P}(P_0 B_0 L_{0,s}^{-1} \geq P_j B_j L_{min,j}^{-1}) \label{Aeq7} \\
&=\left (\prod_{j=1}^2 \mathbb{P}\left(P_0 B_0 L_{0,s}^{-1} \geq P_j B_j L_{j}^{-1}\right)\right) \nonumber \\
&=\hspace{-0.2cm} \int_{\eta_s H^{\alpha_s}}^\infty \hspace{-0.5cm}\mathcal{P}_{s}\left(\left(\frac{l_{0,s}}{\eta_s}\right)^{2/\alpha_s}\hspace{-0.6cm}-H^2\right)  \hspace{-0.1cm}f_{L_{0,s}}(l_{0,s}) \prod_{j=1}^2 \bar{F}_{L_j}\!\!\left(\frac{P_j B_j}{P_0 B_0} l_{0,s}\!\!\right) \hspace{-0.1cm} dl_{0,s} \label{Aeq8} \\
&=\hspace{-0.2cm} \int_{\eta_s H^{\alpha_s}}^\infty \hspace{-0.6cm} \mathcal{P}_{s}\left(\left(\frac{l_{0,s}}{\eta_s}\right)^{2/\alpha_s}\hspace{-0.6cm}-H^2\right) \hspace{-0.1cm}f_{L_{0,s}}(l_{0,s}) e^{ -\hspace{-0.1cm}  \sum_{j=1}^2 \hspace{-0.05cm} \Lambda_j\left(\left[0,\frac{P_j B_j}{P_0 B_0} l_{0,s}\right)\right)} \hspace{-0.1cm} dl_{0,s} \label{app:Lemma3_0}
\end{align}
where (\ref{Aeq7}) follows from the definition of association probability, (\ref{Aeq8}) follows from the fact that there is only one UAV in the $0^{\thh}$ tier which can be a LOS or a NLOS UAV, and CCDF of $L_j$ is formulated as a result of the probability expression and, (\ref{app:Lemma3_0}) follows from the definition of the CCDF of the path-loss.

Association probability with a $1^{\st}$ tier LOS/NLOS UAV can be computed as follows:
\begin{align}
&\mathcal{A}_{1,s}=\mathbb{P}(P_1 B_1 L_{1,s}^{-1} \geq P_j B_j L_{min,j}^{-1}) \mathbb{P} (L_{1,s^{\prime}}>L_{1,s}) \label{Aeq4} \\
&=\left (\prod_{j=0,j \neq 1}^2 \mathbb{P}\left(P_1 B_1 L_{1,s}^{-1} \geq P_j B_j L_{j}^{-1}\right)\right) \mathbb{P} (L_{1,s^{\prime}}>L_{1,s}) \nonumber \\
&=\int_{\eta_s H^{\alpha_s}}^\infty \!\!\!\prod_{j=0,j \neq 1}^2 \!\!\!\!\!\bar{F}_{L_j}\!\!\left(\frac{P_j B_j}{P_1 B_1} l_{1,s}\!\!\right) \!e^{-\Lambda_{1,s^{\prime}}([0,l_{1,s}))} f_{L_{1,s}}(l_{1,s})dl_{1,s} \label{Aeq5} \\
&=\int_{\eta_s H^{\alpha_s}}^\infty \bar{F}_{L_0}\!\!\left(\frac{P_0 B_0}{P_1 B_1} l_{1,s}\right) e^{- \Lambda_2\left(\left[0,\frac{P_2 B_2}{P_1 B_1} l_{1,s}\right)\right)} e^{-\Lambda_{1,s^{\prime}}([0,l_{1,s}))} \label{Aeq6} \\
& \hspace{1cm} \times \Lambda_{1,s}^{\prime}([0,l_{1,s})) e^{-\Lambda_{1,s}\left(\left[0,l_{1,s}\right)\right)}dl_{1,s} \nonumber \\
&=\int_{\eta_s H^{\alpha_s}}^\infty \Lambda_{1,s}^{\prime}\left(\left[0,l_{1,s}\right)\right) \bar{F}_{L_0}\!\!\left(\frac{P_0 B_0}{P_1 B_1} l_{1,s}\right) \nonumber \\
&\hspace{1cm} \times e^{-\sum_{j=1}^2 \Lambda_j\left(\left[0,\frac{P_j B_j}{P_1 B_1} l_{1,s}\right)\right)} dl_{1,s}, \label{app:Lemma3_1}
\end{align}
where $s, s' \in \{\LOS,\NLOS\}$, and $s \neq s'$. (\ref{Aeq4}) follows from the definition of association probability, in (\ref{Aeq5}), CCDF of $L_j$ is formulated as a result of the probability expression, and similarly $\mathbb{P} (L_{1,s^{\prime}}>L_{1,s})=\bar{F}_{L_{1,s^{\prime}}}(l_{1,s})=e^{-\Lambda_{1,s^{\prime}}([0,l_{1,s}))}$; (\ref{Aeq6}) follows from the definition of the CCDF of the path-loss, and by plugging the PDF of the path-loss  $L_{1,s}$; and (\ref{app:Lemma3_1}) follows from the fact that $\Lambda_{1,s}([0,l_{1,s}))+\Lambda_{1,s^{\prime}}([0,l_{1,s}))=\Lambda_{1}([0,l_{1,s}))$. Since the minimum distance between UEs and UAVs is equal to $H$, integration starts from $l_{k,s}=\eta_s H^{\alpha_s}$. Association probability with a $2^{\nd}$ tier BS can be obtained following the similar steps. Note that, since the minimum distance between the typical UE and a ground BS is equal to 0, integration starts from 0.

\subsection{Proof of Theorem 1}
\label{Proof of Theorem 1}
Given that the UE is associated with a UAV in $k=\{0,1\}$, the conditional coverage probability $\PC_{k,s}(\Gamma_k)$ can be computed as follows
\begin{align}
&\PC_{k,s}(\Gamma_k)=\mathbb{P}(\SINR_{k,s}>\Gamma_k) \nonumber \\
&=\mathbb{P}\left (\frac{P_kh_{k,0}L_{k,s}^{-1}}{\sigma_k^2+\sum_{j=0}^2 I_{j,k}} >\Gamma_k \right) \nonumber \\
&=\mathbb{P}\left(h_{k,0}>\frac{\Gamma_k L_{k,s}}{P_k}\left(\sigma_k^2+\sum_{j=0}^2 I_{j,k}\right)\right) \nonumber \\
&=  e^{-u\sigma_k^2} \prod_{j=0}^2  \mathcal{L}_{I_{j,k}}(u),  \label{app:Theorem1}
\end{align}
where $u=\frac{\Gamma_k L_{k,s}}{P_k}$, $\mathcal{L}_{I_{j,k}}(u)$ is the Laplace transform of $I_{j,k}$ evaluated at $u$, the last steps follows from $h_{k,0}$ $\sim$ $\exp(1)$, and by noting that Laplace transforms of interference at the UE from different tier UAVs and BSs are independent. $\PC_{2}(\Gamma_2)$ can be obtained using the similar steps. Tools from stochastic geometry can be applied to compute the Laplace transforms. Recall that $0^{\thh}$ is generated by the UAV at the cluster center of the typical UE. When the typical UE is associated with a UAV or a BS in tier-$k$ for $k=1,2$, Laplace transform of the interference from $0^{\thh}$ tier UAV can be obtained as follows:
\begin{align}
&\mathcal{L}_{I_{0,k}}(u)= \mathbb{E}_{I_{0,k}}\left[e^{-uI_{0,k}}\right]\nonumber\\
&=\!\!\!\!\!\sum_{s^{\prime} \in \{\LOS,\NLOS\}}\!\!\!\!\!\mathbb{E}_x\left[\mathbb{E}_{h_{0,0}}\left[\exp\left(-u P_0h_{0,0}x^{-1}\right)|x>\frac{P_0B_0}{P_k B_k}l_{k}  \right] \right] \label{Aeq9} \\
&=\!\!\!\!\!\sum_{s^{\prime} \in \{\LOS,\NLOS\}} \!\!\!\!\!\mathbb{E}_x\left[ \frac{1}{1+uP_0x^{-1}}|x>\frac{P_0B_0}{P_k B_k}l_{k} \right] \label{Aeq10} \\
&=\!\!\!\!\!\sum_{s^{\prime} \in \{\LOS,\NLOS\}} \int_{\mathcal{E}_{0,0}}^{\infty} \frac{1}{1+uP_0x^{-1}} \mathcal{P}_s(x) f_{L_{0,s^{\prime}}}(x) dx
\end{align}
where conditioning in (\ref{Aeq9}) is a result of the fact that interfering $0^{\thh}$ tier UAV lies outside the exclusion disc $\mathcal{E}_{0,0}$ with radius $\frac{P_0B_0}{P_k B_k}l_{k}$, and (\ref{Aeq10}) follows from $h_{0,0}$ $\sim$ $\exp(1)$. Also note that, $\mathcal{L}_{I_{0,k}}(u)$ is equal to one, if the typical UE is associated with $0^{\thh}$ UAV. Laplace transform of the interference from $1^{\st}$ tier UAVs can be calculated as
\begin{align}
&\mathcal{L}_{I_{1,k}}(u)= \mathbb{E}_{I_{1,k}}\left[e^{-uI_{1,k}}\right]\label{Aeq11}\\
&=\!\!\!\!\!\!\prod_{s^{\prime} \in \{\LOS,\NLOS\}}\!\!\!\!\!\!\!\!\! \exp\left(-\!\!\!\int_{\mathcal{E}_{1,0}}^{\infty}\!\!\!\left(1-\mathbb{E}_{h_{1,i}} \left[ e^{-u P_1 h_{1,i} x^{-1} }\right]\right)\Lambda_{1,s^{\prime}}^{\prime}(dx) \right) \nonumber \\
&=\!\!\!\!\!\!\prod_{s^{\prime} \in \{\LOS,\NLOS\}}\!\!\!\!\!\!\!\!\! \exp\left(-\!\!\!\int_{\mathcal{E}_{1,0}}^{\infty}\!\!\!\left(\frac{u P_1 x^{-1}}{1+u P_1 x^{-1}}\right)\Lambda_{1,s^{\prime}}^{\prime}(dx) \right) \label{Aeq12}
\end{align}
where $\Lambda_{1,s^{\prime}}^{\prime}(dx)$ is obtained by differentiating $\Lambda_{1,s^{\prime}}([0,x))$ given in (\ref{intensity_function_1}) with respect to $x$ for $s^{\prime} \in \{\LOS,\NLOS\}$, respectively, interfering $1^{\st}$ tier UAVs lie outside the exclusion disc $\mathcal{E}_{1,0}$ with radius $\frac{P_1B_1}{P_k B_k}l_{k}$, (\ref{Aeq11}) is obtained by computing the probability generating functional (PGFL) of the PPP, and (\ref{Aeq12}) follows from computing the moment generating function (MGF) of the exponentially distributed random variable $h$. Laplace transform of the interference from $2^{\nd}$ tier BSs, $\mathcal{L}_{I_{2,k}}(u)$, can be calculated following the same steps with the calculation of $\mathcal{L}_{I_{1,k}}(u)$. However, note that there are only LOS BSs for $2^{\nd}$ tier. Finally, by inserting (\ref{Association_Prob0}), (\ref{Association_Prob1}), (\ref{Association_Prob2}), (\ref{LT_I0}), (\ref{LT_I1}), (\ref{LT_I2}) into (\ref{CoverageProbability}), coverage probability expression in (\ref{total_SINR_coverage}) can be obtained.


\end{document}